# Importance of correlation effects in hcp iron revealed by a pressure-induced electronic topological transition

K. Glazyrin[1,2], L.V. Pourovskii[3,4], L. Dubrovinsky[1], O. Narygina[5], C. McCammon[1], B. Hewener[6], V. Schünemann[6], J. Wolny[6], K. Muffler[6], A. I. Chumakov[7], W. Crichton[7], M. Hanfland[7], V. Prakapenka[8], F. Tasnádi[9], M. Ekholm[3], , M. Aichhorn[10], V. Vildosola[11], A. V. Ruban[12], M. I. Katsnelson[13], I. A. Abrikosov[9]

[1] *Bayerisches Geoinstitut, Universität Bayreuth, 95440 Bayreuth, Germany,*

[2]*Yale University, 06511 New Haven, CT, USA*

[3]*Swedish e-science Research Center (SeRC), Department of Physics, Chemistry and Biology (IFM), Linköping University, Linköping, Sweden*

[4¶]*Center de Physique Theorique, Ecole Polytechnique, 91128 Palaiseau Cedex, France*

[5]*School of Physics and Astronomy, University of Edinburgh, Edinburgh, UK*

[6]*Technische Universität Kaiserslautern, Kaiserslautern, Germany*

[7]*ESRF, F-38043 Grenoble Cedex, France*

[8]*GeoSoilEnviroCARS, University of Chicago, Argonne National Laboratory, Argonne, IL 60439, USA*

[9]*Department of Physics, Chemistry and Biology (IFM), Linköping University, Linköping, Sweden*

[10]*Institute of theoretical and computational physics, TU Graz, 8010 Graz, Austria*

[11]*Centro Atómico Constituyentes, GIyANN, CNEA, San Martin, Buenos Aires, Comisión Nacional de Investigaciones Científicas y Técnicas, Ciudad de Buenos Aires, Argentina*

[12]*Department of Materials Science and Engineering, Royal Institute of Technology, SE-10044, Stockholm, Sweden*

[13]*Radboud University Nijmegen, Institute for Molecules and Materials, 6525 AJ, Nijmegen, Netherlands*

## Abstract

We discover that *hcp* phases of Fe and $Fe_{0.9}Ni_{0.1}$ undergo an electronic topological transition at pressures of about 40 GPa. This topological change of the Fermi surface manifests itself through anomalous behavior of the Debye sound velocity, *c/a* lattice parameter ratio and Mössbauer center shift observed in our experiments. First-principles simulations within the dynamic mean field approach demonstrate that the transition is induced by many-electron effects. It is absent in one-electron calculations and represents a clear signature of correlation effects in *hcp* Fe.

Iron is the most abundant element on our planet. It is one of the most important technological materials and, at the same time, one of the most challenging elements for the modern theory. As a consequence, the study of iron and iron-based alloys has been a focus of experimental and computational research over the past decades. Recently, investigations of phase relations and physical properties of iron and its alloys at high pressure led to new exciting discoveries including evidence for a body-centred-cubic (bcc) phase of iron-nickel alloy at conditions of the Earth's core [1] and the observation of superconductivity in the high-pressure hexagonal close packed (*hcp*) phase of iron in the pressure range 15-30 GPa and at temperatures below 2 K [2].

While the structural properties of iron and iron-nickel alloys at pressures below 100 GPa are well established [3], their electronic and magnetic properties are still debated. The $\alpha$-phases (*bcc*) of Fe and $Fe_{0.9}Ni_{0.1}$ are ferromagnetic at ambient conditions, but an accurate description of the electronic structure of $\alpha$-Fe and its high-temperature magnetism require a proper treatment of the many-electron effects [4]. The $\gamma$-phases (face-centered cubic, *fcc*) are believed to have complex incommensurate magnetic ground states, which are still not reproduced by theory [5]**.** The importance of correlation effects for the description of the $\alpha$- to $\gamma$-phase transition in Fe at elevated temperature and ambient pressure has been recently underlined [6]. The $\varepsilon$-phases (*hcp*) of Fe and $Fe_{0.9}Ni_{0.1}$ were previously believed to be nonmagnetic [7]; however recent theoretical work showed that a collinear antiferromagnetic state (AFM-II) [8–10] or a more complex

AFM state [11] have lower energy than the nonmagnetic state. Nevertheless, the AFM-II phase could not be resolved in Mössbauer experiments. Moreover, theoretical estimates of the Néel temperature $T_N$ yield a maximum value of ~69 K for *hcp* Fe at the transition pressure (12 GPa), followed by a decrease with increasing pressure [12]. Although nickel atoms are predicted to enhance the magnetic moments on neighboring iron atoms, there is no evidence that ε-$Fe_{0.9}Ni_{0.1}$ is a static antiferromagnet down to at least 11 K at 21 GPa [10], implying that direct comparison is unreliable between static (0 K) *ab initio* calculations for AFM ε-Fe and room temperature experimental data that clearly indicate a paramagnetic phase. It is worth noting that *hcp* Fe becomes superconducting in the same pressure range [2], and that the mechanism of superconductivity is believed to be unconventional [13]. These observations indicate that the physical behavior of *hcp* Fe at moderate pressures below 70 GPa is complex and the role of correlation effects beyond the standard density-functional (DFT) approach in the physics of this material is not well understood.

In order to unravel the evolution of the electronic structure in *hcp* Fe and $Fe_{0.9}Ni_{0.1}$ under pressure we have carried out a combined experimental and theoretical investigation. We have extracted the Debye sound velocity $V_D$ for pure Fe and $Fe_{0.9}Ni_{0.1}$ alloy from nuclear inelastic scattering (NIS) experiments as well as precisely measured the lattice parameter *c/a* ratio and the Mössbauer centre shift in the pressure range from 12 to 70 GPa. All of our results show anomalous behavior at a similar pressure ~ 40 GPa. Our state-of–the-art *ab initio* simulations within the dynamical mean-field theory [14–16] reveal an electronic topological transition (ETT) in the *hcp* phase of iron at pressures of about 30-40 GPa, providing an explanation of the experimentally observed anomalies. The absence of the ETT in conventional one-electron DFT calculations demonstrates that many-body correlation effects determine the Fermi surface topology of paramagnetic *hcp* Fe, and, therefore, essential for the correct description of the complex physical phenomena observed in this material.

Figure 1 summarizes our experimental measurements of the Debye sound velocity $V_D$ for Fe and $Fe_{0.9}Ni_{0.1}$ extracted from NIS experiments (technical details are given in Supplementary Information [17]). The experimental data show a softening of $V_D$ in the pressure range 42-52 GPa. To verify our results we also analyzed the available literature [18–21] and conclude that the same softening of $V_D$ has been observed at pressures of 40-50 GPa. The phenomenon was not given much attention in the previous publications, perhaps due to data scatter and the uncertainties of individual data points.

The softening of the Debye sound velocity in Fig. 1 is weak, so we made further investigations. We measured the lattice parameters of *hcp*-Fe in a diamond anvil cell (DAC) on compression to ~65 GPa in quasi-hydrostatic He pressure transmitting medium at ambient temperature and found an anomaly in *c/a* at about 40 GPa (Fig. 2a), consistent with the pressure at which $V_D$ shows softening. The pressure dependence of the *c/a* ratio in *hcp* Fe has been the subject of several previous experimental studies [22–27] that were mainly focused on much higher pressures. However, a closer inspection of the results by Dewaele *et al.* [24] shows very good agreement with our data (*Fig. S.1.3* [17]). Also, an anomalous behavior of *c/a* was reported at about 50 GPa based on a limited number of data points collected in DAC experiments using a non-hydrostatic (NaCl) pressure-transmitting medium [27].

Mössbauer spectroscopy can also be a powerful method to detect pressure-induced transitions [28]. We performed Mössbauer experiments on pure Fe and $Fe_{0.9}Ni_{0.1}$ up to 60 GPa in a DAC loaded with He as a quasi-hydrostatic pressure transmitting medium, and observed a large anomaly in the center shift variation with pressure at 40-45 GPa (Fig. 2b). Our theoretical calculations demonstrate that the anomaly cannot be explained by changes of the electron density at the nuclei and, correspondingly, of the isomer shift [17]. Therefore, the anomaly must be associated with the second-order Doppler shift [28].

We have shown from three independent experimental methods pressure-induced anomalies in the pressure range 40-50 GPa. We note that X-ray diffraction does not reveal any crystallographic structural change of *hcp*-Fe and $Fe_{0.9}Ni_{0.1}$ at the same conditions [1,29,30], and as discussed above, there is no long range magnetic order in the *hcp* phase of Fe detected by experiments. The observed anomalies must therefore be associated with changes in the electronic state of paramagnetic *hcp*-Fe and $Fe_{0.9}Ni_{0.1}$. To address this question we made a theoretical investigation of the electronic structure of ε-Fe at moderate pressures in the range 12-70 GPa. We employed a state-of-the-art fully self-consistent technique [16] combining full-potential linearized augmented plain-wave (LAPW) band structure method with the dynamical mean-field theory (DMFT) treatment of the on-site Coulomb repulsion between Fe 3*d* states [17]. The DMFT quantum impurity problem was solved using the exact Continuous-time strong-coupling Quantum Monte-Carlo method [31]. A combination of LDA and DMFT has been applied previously to investigate thermodynamic stability [6] and to describe the magnetic properties [4] of paramagnetic *bcc* Fe at ambient pressure, which justifies the choice of method for this work.

The LDA+DMFT Fermi surfaces and ***k***-resolved spectral functions for two different volumes are shown in Fig. 3. The *hcp* phase of Fe is predicted to be weakly correlated, with the average mass enhancement decreasing from 1.43 at 16 GPa to 1.25 at 69 GPa, indicating a reduced correlation strength at smaller volumes. Sharp bands in the vicinity of the Fermi level $\varepsilon_F$ and a noticeable shift of bands toward $\varepsilon_F$ compared to the LDA picture (Fig. 3 (e) and (f)) are the usual features of a Fermi liquid. Most interestingly, the hole-like bands at the **Γ** and **L** points visible at smaller volume are found *below* $\varepsilon_F$ at V=10.4 Å$^3$/at. Hence, the DMFT calculations show that the topology of the Fermi surface changes under compression. Indeed a comparison of Figs. 3 (a) and (b) shows that hole pockets appears at **Γ** and **L** with decreasing volume, and therefore *hcp* Fe undergoes an *electronic topological transition* [32] under applied pressure. The actual ETT takes place at P~30 GPa. It is remarkable that the observed ETT is absent in the LDA calculations; it appears only upon inclusion of correlation effects.

The effects of ETT on the lattice properties of metals within the one-electron approximation are well understood [33]. The elastic moduli $C_{ii}$ calculated at the condition of constant particle number at the deformation contains the contribution

$$\delta C_{ii} = -\frac{1}{V_0}\sum_{\lambda}\left(\frac{\partial \xi_{\lambda}}{\partial u_i}\right)^2 \delta(\xi_{\lambda}), \qquad (1)$$

where $\xi_{\lambda} = \varepsilon_{\lambda} - \varepsilon_F$ , and $\varepsilon_{\lambda}$ denotes the single-particle energies. $\xi_{\lambda}$ is singular near the ETT, and this singular contribution has the same singularity as $-N(\varepsilon_F)$. This means, in particular, that the peculiarity in the Debye sound velocity is $\Delta V_D \sim -\delta N(E_F)$, where $\delta N(E_F)$ is the change in the density of states (DOS) at the Fermi level due to ETT. In the case of an appearance of a new hole pocket below the critical volume $V_{ETT}$ the change in DOS is $\delta N(E_F) \sim (V_{ETT}-V)^{1/2}$, hence the one-electron theory predicts the existence of square-root-down-shaped peculiarity at the ETT. Our DMFT calculations show that in the case of *hcp*-Fe at moderate compression one should use the Fermi-liquid theory of ETT [34]. In this case many-electron effects cause the singularity of the thermodynamic potential $\Omega$ at ETT to be two-sided. Still the leading term is a square root in $\Delta V_D$ on one side of the transition, while the peculiarity on the other side of the transition is one power weaker.

The Debye temperature $\theta_D$ also has a singularity as $-N(\varepsilon_F)$, and lattice heat capacity at low temperature $T << \theta_D$ has the same singularity as the electron heat capacity. The thermal expansion coefficient proportional to the derivative of $\theta_D$ with respect to deformation has a stronger singularity at these temperatures, like $\frac{\partial N(\varepsilon_F)}{\partial \varepsilon_F}$, that is divergent at the point of ETT (e.g., [35]). It is important to stress, however, that the Debye model is qualitatively incorrect in the situation of ETT. Strong anomalies of the phonon spectra in the harmonic approximation occur in a relatively small part of the Brillouin zone near the Γ point and the average phonon frequency over the whole Brillouin zone, which is relevant for thermodynamics at $T \approx \theta_D$, is weaker by a factor of $\varepsilon_F - \varepsilon_c$, where $\varepsilon_c$ is the Van Hove singularity energy [36]. However, if we take into account quasiharmonic and anharmonic effects, i.e., the temperature dependence of phonon frequencies due to thermal expansion and phonon-phonon interactions, the singularities again enhance and become like $N(\varepsilon_F)$ in average phonon frequencies and like $\frac{\partial N(\varepsilon_F)}{\partial \varepsilon_F}$ in the elastic moduli [36].

For *hcp* metals ETTs have been associated with anomalies in the lattice parameter ratio *c/a* in the vicinity of the transition [37–40]. The dependence of lattice constants on the external parameters is less singular than $C_{ii}$ since they are related to the first derivatives of the thermodynamic potential, while $C_{ii}$ are related to the second derivatives. This means that the anomaly in the *c/a* ratio at zero temperature should be hardly visible but at finite (and sufficiently high) temperatures it is proportional to $N(\varepsilon_F)$ via the anomaly of the thermal expansion coefficient, discussed above. The same is true for the second-order Doppler shifts of the Mössbauer spectra related to the heat capacity and, thus, with the average phonon frequencies over the Brillouin zone. Thus, the theory of ETT provides a convincing explanation of the experimentally observed anomalies of the sound velocity, *c/a* ratio and center shift at 40-45 GPa.

To conclude, we observe the electronic isostructural transition of *hcp* Fe and $Fe_{0.9}Ni_{0.1}$ at a pressure of ~40 GPa. The presence of the transition is confirmed by three independent experimental approaches – nuclear inelastic scattering, *c/a* ratio measurement, and Mössbauer center shift determination. The theoretical calculations carried out by means of state-of-the-art *ab initio* methods explain the anomalies in terms of a change of the Fermi surface topology, a so-called electronic topological transition. The existence of the ETT in many-body calculations and its absence in one-electron calculations is a clear signature of correlation effects in the paramagnetic phase of *hcp* Fe. Therefore, advanced approaches beyond the density functional theory are needed to understand the complex physics of this material. Our results also point out to possible importance of many-body effects in other itinerant metallic systems at high-pressure conditions.

We are grateful to Prof. A. Georges for useful discussion. The funding provided by Swedish e-science Research Centre (SeRC), the Swedish Research Council via grant 621-2011-4426, and the Swedish Foundation for Strategic Research (SSF) programs SRL grant 10-0026 and “Multifilms”,as well as financial support from German Science Foundation (DFG) and German Ministry for Education and Research (BMBF) are acknowledged. Calculations have been performed at the Swedish National Infrastructure for Computing (SNIC). We acknowledge the European Synchrotron Radiation Facility for provision of synchrotron radiation facilities on beamlines ID09a and ID18.

**Figure captions:**

**Figure 1 (color online)**

Debye sound velocity $V_D$ as a function of pressure for pure iron (filled black squares, -1-) and $Fe_{0.9}Ni_{0.1}$ alloy (open triangles,-2-).. The upper axis shows the density scale. Also shown are literature data on sound velocities obtained with NIS (open circles [18],-4- and half-filled circles [19],-5-), and impulsive stimulated light scattering (ISLS) measurements [20] (circles with crosses, -6-) for pure ε-Fe, as well as NIS data [21] for ε-$Fe_{0.92}Ni_{0.08}$ (blue open squares,-3-). Experimental data presented in the figure show the softening of $V_D$ in a pressure region of 42-52 GPa.

**Figure 2**

Experimental pressure dependence of (a) *hcp* phase lattice parameter c/a ratio and (b) the Mössbauer centre shift based on several experimental datasets for pure iron (red circles) and for $Fe_{0.9}Ni_{0.1}$ alloy (blue circles). The centre shift values are given relative to pure *bcc* iron. Straight grey lines in (a) are guides for the eye.

**Figure 3.**

The LDA+DMFT ***k***-resolved spectral function *A(**k**,E)* ( in $V_{at}$/eV, where $V_{at}$ is the volume per atom) of *hcp* Fe at volumes of 8.9 Å$^3$/at (*a*) and 10.4 Å$^3$/at (*b*) corresponding to pressures of 69 and 15.4 GPa, respectively. The energy zero is taken at the Fermi level. The hole-like bands at the **Γ** and **L** points at volume 8.9 Å$^3$/at (indicated by the white arrows) are *below* $E_F$ at V=10.4 Å$^3$/at. The corresponding LDA band structures are shown in *e* and *d*, respectively. In (e) and (f) the corresponding LDA+DMFT Fermi surfaces are shown for the same volumes. The full Fermi surface is plotted on the left-hand side and its cut along the Γ-M direction is displayed in the right-hand side. Changes of the FS topology around the L and Γ points are clearly seen.

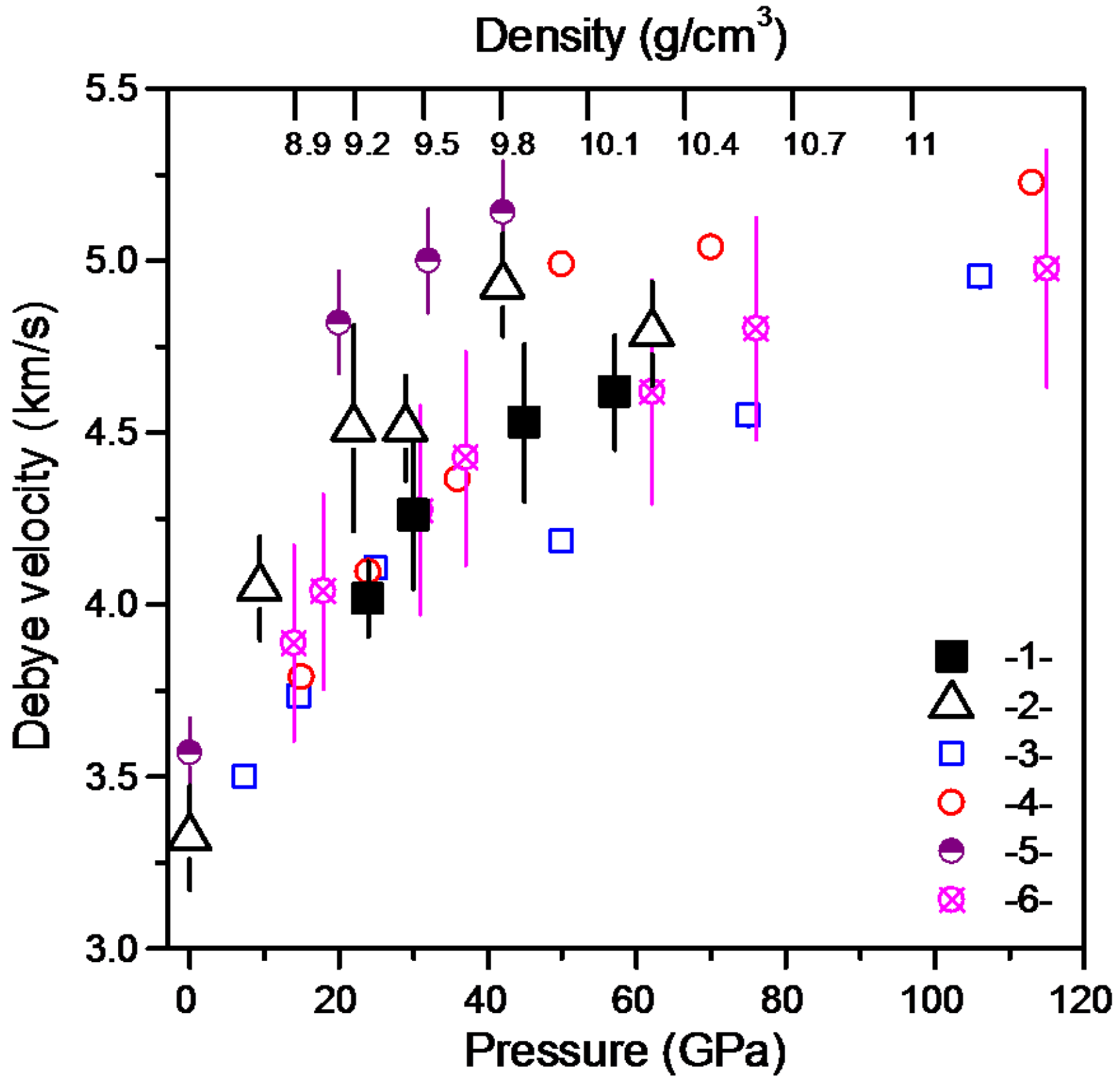

**FIG. 1 (color online)**

314

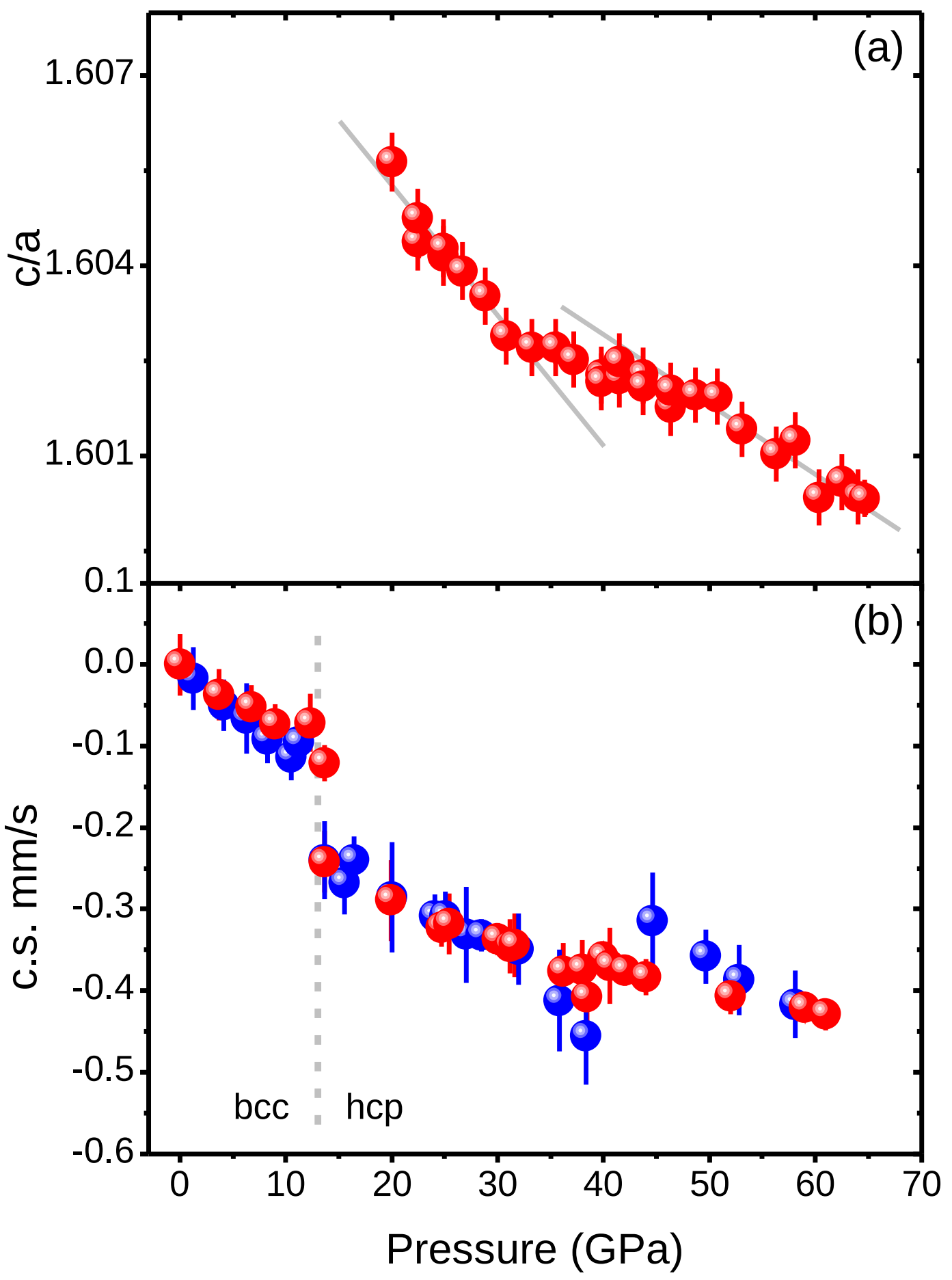

**FIG. 2**

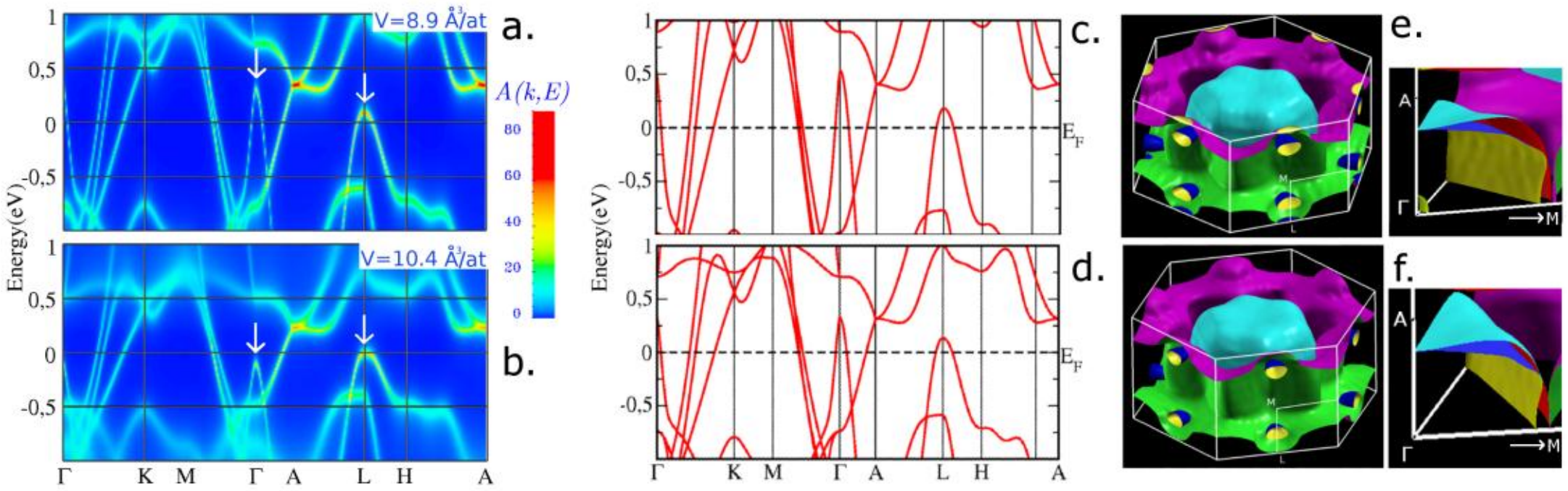
V=8.9 Å³/at
a.
V=10.4 Å³/at
b.
A(k,E)
Energy(eV)
Γ K M Γ A L H A
c.
d.
E_F
e.
f.

**FIG. 3.**

SUPPLEMENTARY INFORMATION

## 1. *Experimental details*

For the preparation of the iron sample we used enriched $^{57}Fe$ of 99.99% purity. The $Fe_{0.9}Ni_{0.1}$ samples were prepared by mixing appropriate amounts of $^{57}Fe$ enriched iron and nickel powder of 99.999% purity. The mixture was compressed to 2 GPa in a piston-cylinder apparatus, heated above the melting point and subsequently quenched.

To achieve high pressures in different experiments we used two types of diamond anvil cells: one piston-cylinder conventional cell for Mossbauer experiments, with He gas for quasi-hydrostatic conditions loaded into a chamber of Re gasket, and for nuclear inelastic x-ray scattering (NIS) another panoramic cell produced in Bayerisches Geoinstitut. Special types of new cells slits make it possible to measure nuclear forward scattering (NFS) or X-Ray diffraction patterns at the same time as NIS, and, due to the specific geometry of NIS cells, Be was chosen as a material for the gaskets. We used LiF as a pressure medium for the NIS cells. In both experiments we used diamonds with 250µm culet size and small ruby chips to measure the pressure inside gasket chambers. The prepared gasket chambers were 60µm·20µm and 100µm·30µm (diameter·height) for NIS and Mossbauer experiments, respectively.

NIS studies were carried out at the beam line ID18 [S1] (ESRF, Grenoble); the details of the experiment are described in Lübbers et al. [S2] and references therein. Data analysis was performed using program DOS-2.1 according to the calculation procedure described in Kohn and Chumakov [S3], and more details of the calculation procedure is described in Sturhahn [S4] and references therein.

To extract the partial density of states for iron (PDOS) using NIS we employed the following procedure. First, we collected absorption spectra of the sample enriched with $^{57}Fe$ as a function of the energy of incident radiation (NIS spectra Fig S1.1). After extracting the elastic contribution to the NIS spectra, we used DOS-2.1 software to extract the PDOS. Next, we calculated Debye sound velocities ($V_D$) using the relation [S5]:

$$\rho \cdot {V_D}^3 = \frac{4 \cdot \pi \cdot \tilde{m}}{h^3} \cdot \lim_{E \to 0} \frac{E^2}{D(E)},$$

where $\rho$ is the density of the material; $\tilde{m}$ is the mass of the nuclear resonant isotope ($^{57}Fe$); $h$ is Planck's constant; $D(E)$ is the iron PDOS and $E$ is energy.

We determined the equation of state for the *hcp* phase of Fe and $Fe_{0.9}Ni_{0.1}$ (the variation of $\rho$ with pressure) and the bulk modulus in a separate synchrotron X-ray diffraction experiment on beam lines ID09a (ESRF, Grenoble) and IDD-13 (APS, Argonne). The parameters derived from a fit to the Birch-Murnaghan equation of state for *hcp*-$Fe_{0.9}Ni_{0.1}$ are bulk modulus K=159(3) GPa, K'=4.78(9) and $V_0$=6.77(1) $cm^3$/mol, and for *hcp*-Fe EOS these parameters have values indistinguishable from the values reported in previous studies [S6,S7]. We collected a set of X-ray diffraction data for pure Fe in a compression run up to ~65 GPa (quasi-hydrostatic pressure medium - He). We used the ruby fluorescence method for pressure determination [S8] and Fullprof [S9] and Unitcell [S10] programs to extract the *c* and *a* lattice parameters for *hcp*-Fe. There is a remarkable agreement between values calculated by these programs. A representative X-diffraction pattern from the second run is shown in Fig. S1.2. The results of our runs are compared in Fig. S1.3 with data from the literature [S11,S12].

$^{57}Fe$ Mössbauer spectra (MS, Fig S1.4) were recorded at room temperature in transmission mode on a constant acceleration Mössbauer spectrometer using a nominal 370 MBq $^{57}Co$ high specific activity source in a 12 μm Rh matrix (point source). The velocity scale was calibrated relative to 25 μm α-Fe foil. Mössbauer spectra were fitted to Lorentzian lineshapes using the commercially available fitting program NORMOS written by R.A. Brand (distributed by Wissenschaftliche Elektronik GmbH, Germany). Collection time for each spectrum varied from 24 to 48 h.

In our experiments the *bcc* α-phase transforms to the *hcp* ε-phase at 10-13 GPa, in good agreement with literature data [S13,S14]. The center shift in hcp-Fe and hcp-$Fe_{0.9}Ni_{0.1}$ gradually decreases with pressure, up to 40-45 GPa. In this pressure range we observe an abrupt increase of the center shift by ~0.15 mm/s in $Fe_{0.9}Ni_{0.1}$ and a smaller value of ~0.05 mm/s in pure iron (Fig. 2, main text). Upon further compression to 60 GPa we see no sign of further irregular behavior. The discontinuity was observed in several independent DAC loadings. Previous X-ray diffraction measurements have not revealed any structural transformation in the pressure range 13-60 GPa in both *hcp* Fe and $Fe_{0.9}Ni_{0.1}$ [S15,S16].

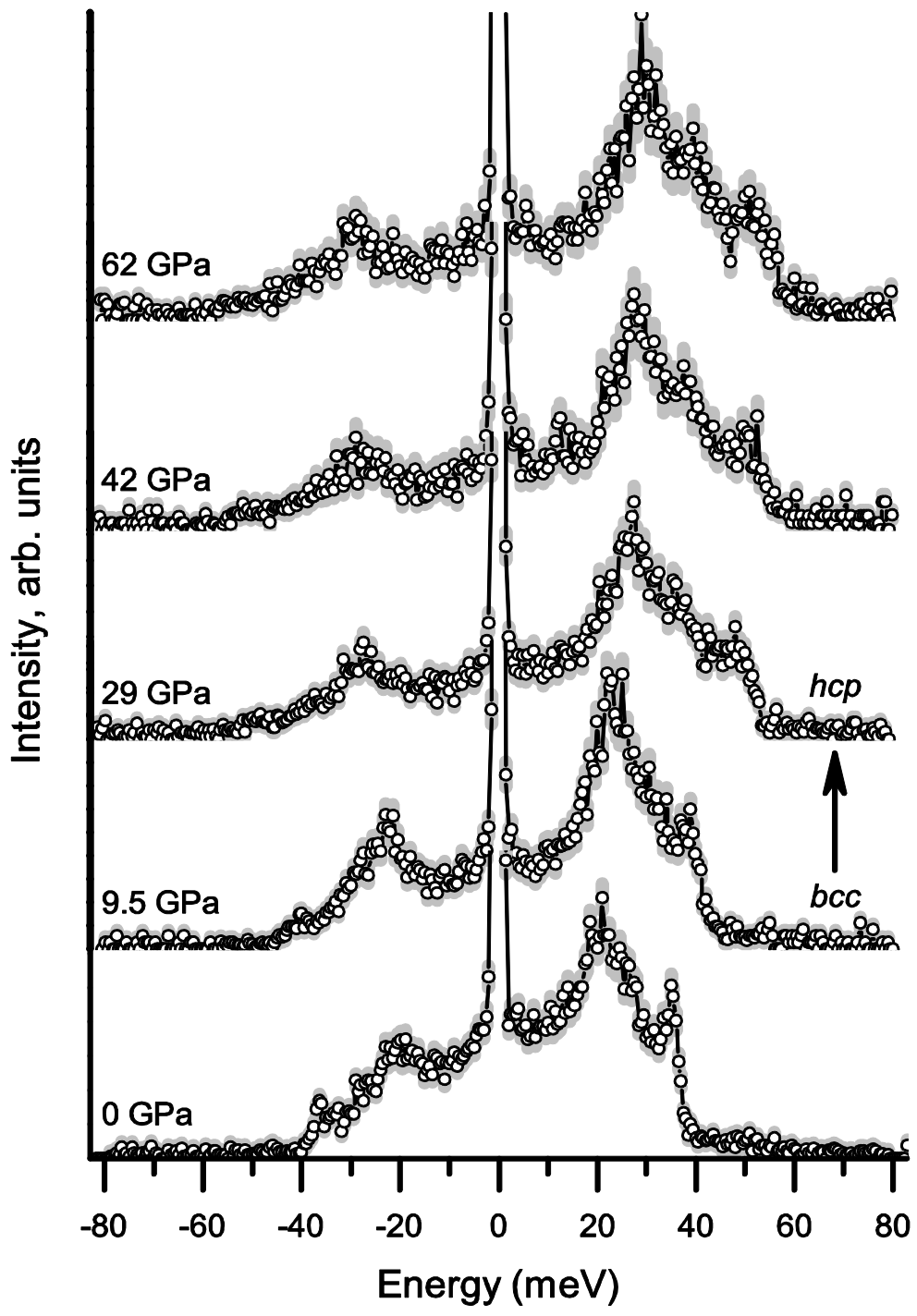

**Fig. S1.1** Nuclear inelastic scattering absorption spectra at various pressures measured on $Fe_{0.9}Ni_{0.1}$.

378

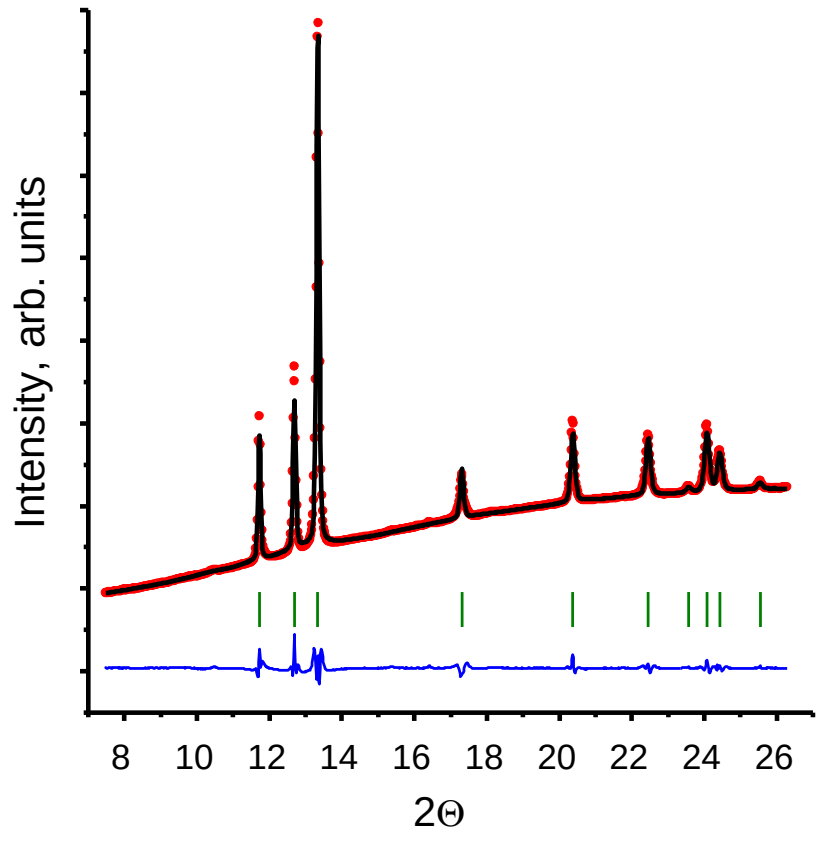

**Fig. S1.2** X-ray diffraction pattern of pure Fe obtained at 64.7(1) GPa. The red data points are the experimental data. The black line, the blue line and the green vertical lines represent a Le Bail fit of data (Fullprof), difference between the data and the model, and positions of *hcp*-Fe diffraction lines, respectively.

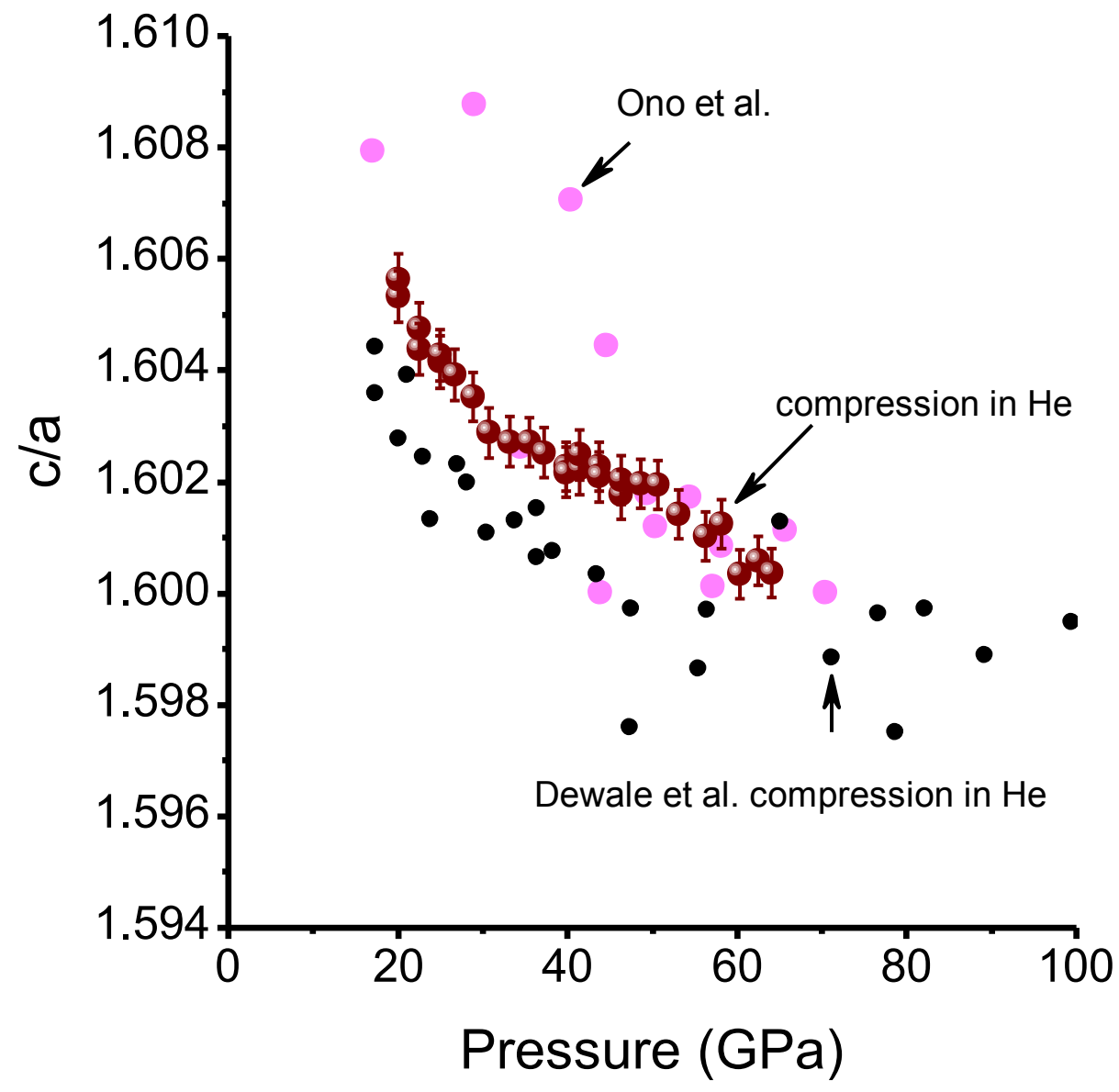

**Fig S1.3** Variation of *c/a* lattice parameter ratio as a function of pressure. Dark red points correspond to our data (He pressure medium), and pink points indicate those obtained by Ono et al. [S11] (NaCl pressure medium, experimental uncertainties unknown) collected during compression runs. The Dewaele *et al.* [S12] data are shown by black closed circles.

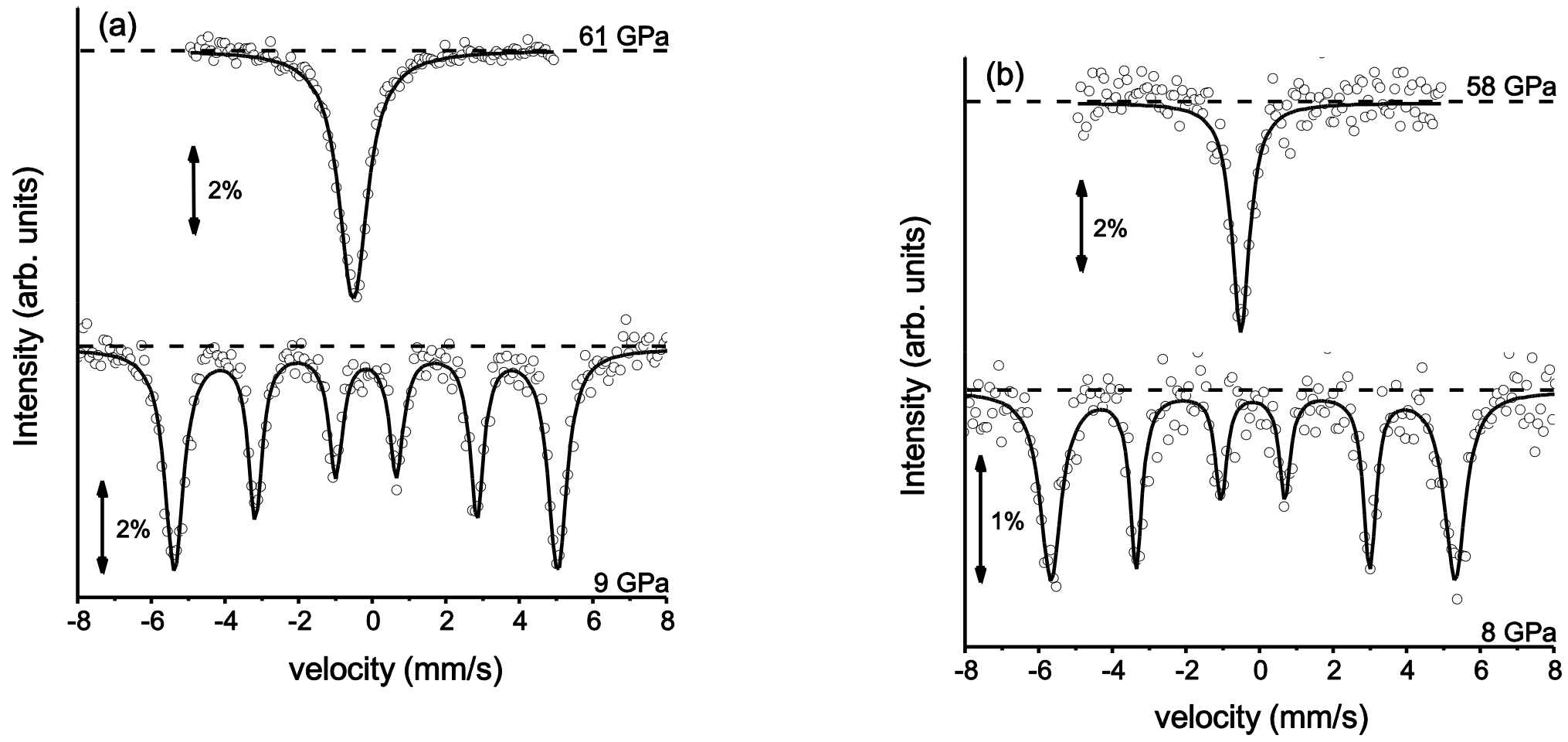

**Fig. S1.4** Representative Mössbauer spectra of (a) pure Fe and (b) $Fe_{0.9}Ni_{0.1}$ samples at the indicated pressures. Solid lines represent model fits.

## *2 Theoretical Simulations*

### *2.1 Local-density approximation+dynamical mean-field theory (LDA+DMFT) approach*

Our LDA+DMFT calculations have been performed with a full-potential fully-self consistent implementation [S17,S18] combining the dynamical-mean-field-theory [S19–S21] and a highly precise full-potential linearized augmented plane wave (FLAPW) electronic structure technique [S22]. The DMFT quantum impurity problem was solved by the continuous-time quantum Monte Carlo (CTQMC) [S23] method. At the beginning of each LDA+DMFT iteration, the eigenvalues and eigenvectors of the Kohn-Sham problem were obtained by the FLAPW code. Wannier-like functions for the Fe-3*d* shell were then constructed by projecting local orbitals onto a set of FLAPW Bloch states located within the energy window from -9.5 to 9.2 eV (details of the projection procedure can be found elsewhere [S17]). We then introduced a local Coulomb interaction acting between those Wannier orbitals and solved the resulting many-body problem within the DMFT framework. After completing the DMFT cycle we calculated the resulting density matrix in the Bloch states' basis, which was then used to recalculate the charge density and update the Kohn-Sham potential at the next iteration [S18].

We have employed the density-density form of the local Coulomb interaction in the spherical approximation, in which case the interaction matrix is fully specified by the parameters U=3.4 eV and J=0.9 eV, these values of U (or Slater parameter $F_0$) and J were obtained in recent constrained random-phase approximation calculations for *bcc* Fe [S24] (note that the parameter U reported in Ref. 22 is the on-site repulsion $U_{mm}^{\uparrow\downarrow}$ between electrons of opposite spins located on the same orbital as well as $J = \frac{1}{N(N-1)} \sum_{m \neq m'}^{N} \left( U_{mm'}^{\uparrow\downarrow} - U_{mm'}^{\uparrow\uparrow} \right)$ , which, after conversion in accordance with formulae provided in Ref. 15, gives the values used in the present work). The quantum impurity problem has been solved with 3 million CTQMC cycles with each cycle consisting of 200 CTQMC moves following by a measurement. The *hcp* phase under consideration is a metal with a rather uniform occupancy of the orbitals within the Fe 3*d* shell. Therefore, we employed the around mean-field form [S25] for the double-counting correction term.

The CTQMC solver computes the Green's function on the imaginary-time axis so an analytic continuation is needed in order to obtain results on the real axis. For that we use a stochastic version of the Maximum Entropy method [S26] with which we calculate the retarded lattice Green's function, $G_{\nu\nu}(k,\omega^{+})$, where $\nu$ labels all the Bloch bands considered. We finally obtain the spectral function as

follows:

$$A(k,\omega) = -\frac{1}{\pi}\mathrm{Im}\left[\sum_{\nu} G_{\nu\nu}(k,\omega^{+})\right].$$

*2.2 Isomer shift calculations in Fe*

The Mössbauer centre shift of the absorption line, $\Delta E^{C}$, is composed of two contributions:

$$\Delta E^{C} = \delta^{IS} + \delta^{SOD},$$

called the isomer shift (IS) and the second-order Doppler shift (SOD), respectively. The latter is a relativistic Doppler shift of the photon frequency due to the relative velocity, $v$, of the source and absorber. Due to the thermal motion of the ions, the corresponding shift of the absorption line $E_{\gamma}$ depends on $v$ to second order as:

$$\delta^{SOD} = \frac{\langle v^{2} \rangle}{2c^{2}} E_{\gamma},$$

where $c$ is the speed of light. The isomer shift, which is due to the different chemical environments of the source and the absorber, can in turn be calculated as:

$$\delta^{IS} = -\alpha[\rho_{a}(0) - \rho_{s}(0)],$$

where $\rho_{a}(0)$ and $\rho_{s}(0)$ are the electron densities in the absorber and the source materials, respectively, evaluated at the nucleus. $\alpha$ is an empirical nuclear calibration constant to which we assigned the value 0.27 $a_{0}^{3}$ mm/s, which is within the common range reported in the literature [S27–S29]. We evaluated $\rho_a(0)$ and $\rho_s(0)$ from *ab initio* calculations, based on the local density approximation (LDA) and local density approximation + dynamical mean-field theory (LDA+DMFT) [S17–S21].

The pure LDA calculations were performed using two different computational methods: exact muffin-tin orbitals (EMTOs) [S30] and FPLAPW [S22]. In EMTO calculations, we treated 3d and 4d electron states as valence, and the core states were recalculated in each iteration of the self-consistency cycle. The number of **k**-points in the numerical integrals over the Brillouin zone was converged with respect to the total energy to within 0.1 meV/atom. In FPLAPW calculations, electronic states were separated into core- and valence-states at -6.0 Ry. The k-points were chosen from a mesh of 10 000 points in the Brillouin zone, and we set the $R_{MT} * K_{\max}$ parameter to 8.0.

In DMFT+LDA calculations, the FPLAPW method was used for the LDA part, as described in Sec. 2.1. We used the experimental lattice constants at the corresponding pressures, which gives a highly accurate description of electronic structure at fixed volume [S31]. The source system was taken to be body-centered cubic (*bcc*) Fe at ambient pressure, in order to emulate the experimental set-up.

Since only *s*-orbitals have finite probability density at the origin, it is reasonable to assume that $\delta^{IS}$ depends mainly on the difference in the *s*-electron charge density at the absorber and emitter nuclei. In the electronic structure computational methods employed in this work, the nucleus is assumed to be point-like and to be situated at the origin. As the *s*-electron wave function diverges in this limit, we have modeled the nucleus as a uniformly charged sphere of radius $R_n$, where:

$$R_n = CA^{1/3}$$

for an isotope of atomic mass *A*, and *C* = 1.25 fm [S32]. Rather than using the value $\rho = \rho(r = R_n)$ in Eq. (3), we take the average density:

$$\langle \rho \rangle = \frac{4\pi}{V_n} \int_{R_0}^{R_n} \rho(r) r^2 dr$$

evaluated from the radial point $R_0$, which is chosen to lie very close to the origin.

We have performed calculations assuming a paramagnetic state. For the LSDA calculations it was simulated by means of disordered local moments (DLM) [S33]. The DLM state was treated using the coherent potential approximation (CPA) within the EMTO framework [S30]. The magnitudes of the local moments were fixed to 0 (corresponding to non-magnetic calculations), 0.5, 1.0, and 1.5 $\mu_B$ in the entire volume range. We also used values determined self-consistently as a function of volume using the method developed in Ref. [S34], which includes longitudinal spin fluctuations in a single-site mean-field approximation. Within the longitudinal spin fluctuation theory only the so-called *on-site* longitudinal spin fluctuation energy, *E(m)*, is calculated by a constrained DFT approach with electronic excitations at temperature *T*. Then, the average value of the magnetic moment, $<m>$, is determined using the corresponding partition function, $Z_m$, assuming that there is full coupling of the transverse and longitudinal degrees of freedom:

$$\langle m \rangle = \frac{1}{Z_m} \int m^3 \exp(-E(m)/k_B T) dm$$
$$Z_m = \int m^2 \exp\left(-E(m)/k_B T\right) dm$$

the resulting average magnetic moments are shown in Fig. S2.2.1 as a function of volume at 300 K.

Fig. S2.2.2 shows the calculated isomer shift as a function of unit cell volume in Fe, which is also compared to experimental measurements of the center shift. Using LDA, the trends of the obtained isomer shift follow the experimental center shift, and the shift due to *bcc-hcp* transition is also seen. However, the anomaly observed in the experimental center shift around the unit cell volume of 65 a.u.$^3$/atom is not seen in the calculated isomer shift, regardless of the assumed magnetic state.

Within the LDA+DMFT technique we obtain a slight offset with respect to the LDA data. This offset is far greater than the statistical fluctuations of the isomer shift during the computational iterations, which are indicated by error bars in Figure S2.2.2. The error bars correspond to +/- 2 standard deviations of the obtained isomer shift during the final ten iterations. In any case, the center shift anomaly is not seen in this case either. Therefore, it should be attributed to the peculiarity of the second-order Doppler shift [S35].

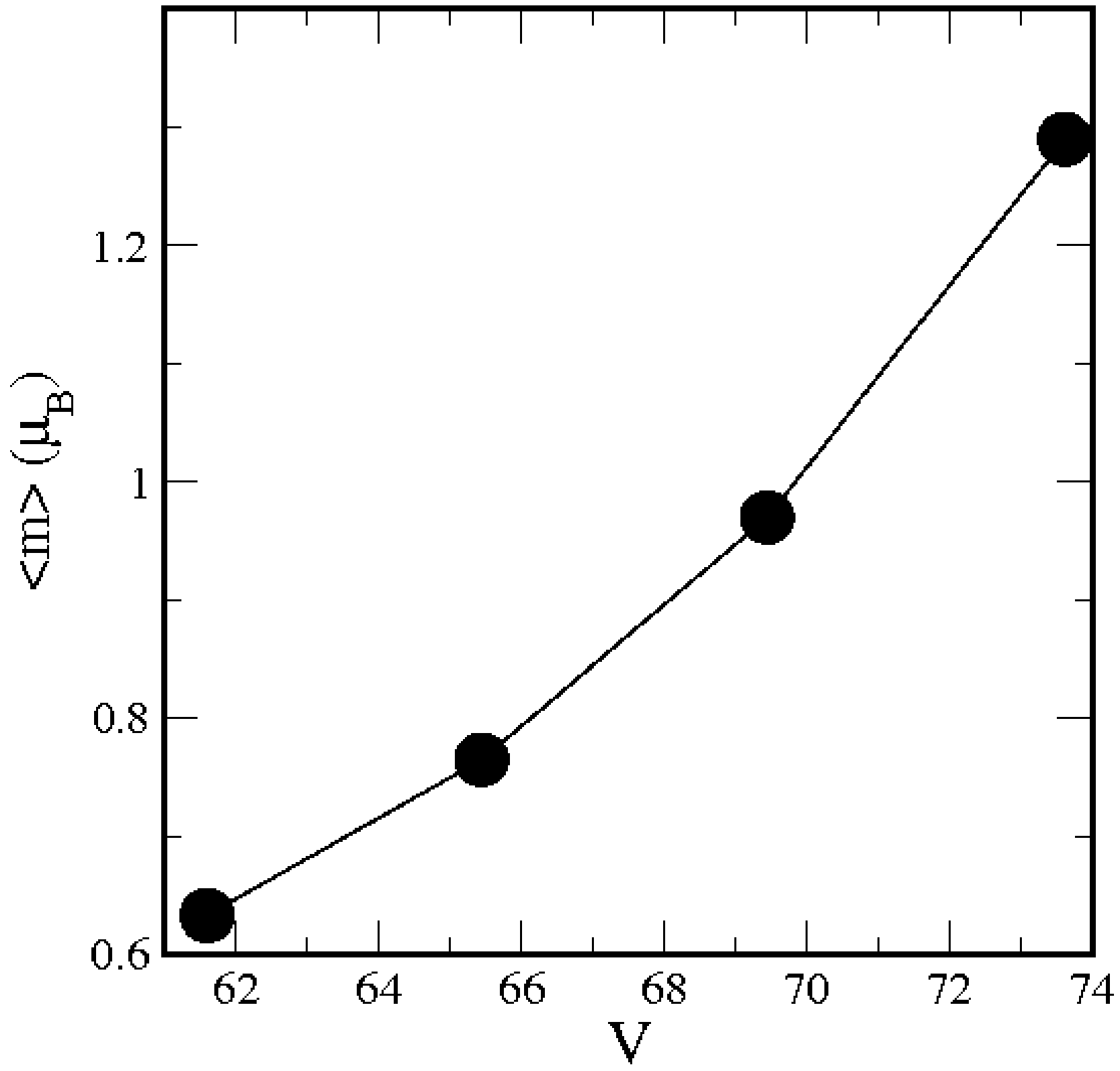

**Fig. S2.2.1** Average magnetic moment in the paramagnetic state of hcp Fe at T=300K as a function of volume per atom (in Bohr$^3$), as calculated by means of the longitudinal spin fluctuation theory.

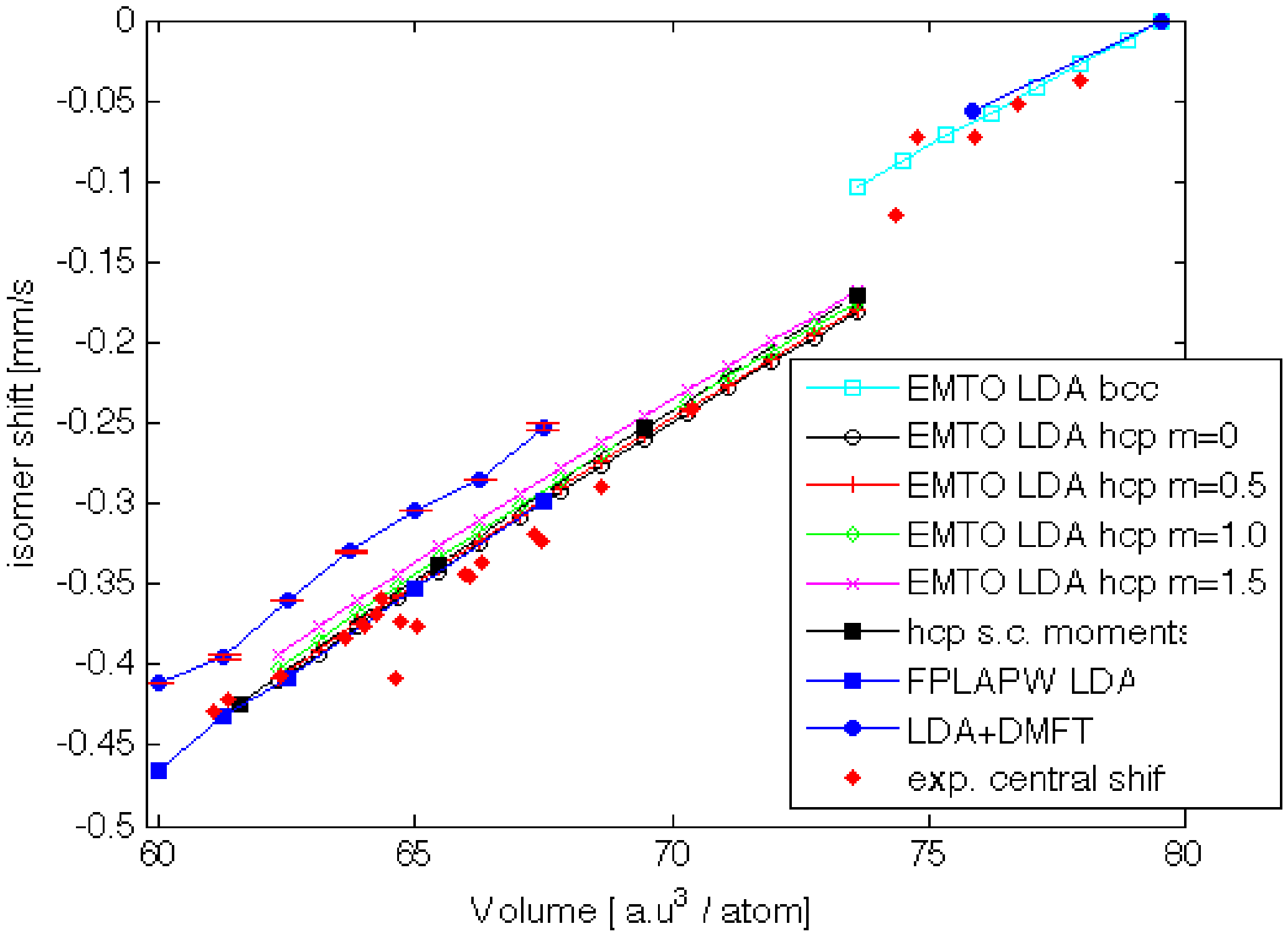

**Fig. S2.2.2** Calculated Mössbauer isomer shifts for pure Fe obtained within different approximations as a function of volume, compared to experimental measurements of the centre shift (red solid diamonds). Results for the *bcc* phase (open cyan squares) are shown as obtained using the EMTO-LDA method. Calculations for the *hcp* phase include a non-magnetic state within the LDA based on the EMTO (open black circles) and FPLAPW methods (blue solid squares), as well as paramagnetic states based on the EMTO method with local magnetic moments 0.5 $\mu_B$ (red plus-signs), 1.0 $\mu_B$ (green open diamonds), and 1.5 $\mu_B$ (pink crosses) and self-consistent moments (Fig. S2.2.1) obtained by spin-fluctuation theory (black squares). We also present results obtained using the LDA+DMFT method (blue filled circles). The error bars correspond to +/- 2 standard deviations of the obtained isomer shift in the LDA+DMFT iterations. All shifts are relative to a reference point at ambient pressure, which was recalculated within all above methods for a correct evaluation of the shifts.